\documentclass[11pt,aps,prd,a4paper,showpacs,showkeys,superscriptaddress, preprintnumbers,floatfix,nofootinbib,notitlepage]{revtex4-1}

\usepackage[normalem]{ulem}
\usepackage{amstext}
\usepackage{amssymb}
\usepackage{amsmath}
\usepackage{graphicx}
\graphicspath{{plots/}}
\usepackage{url}
\usepackage{color}
\usepackage{ulem}
\usepackage[utf8]{inputenc}
\pdfoutput=1
\usepackage{textcomp}
\usepackage{booktabs}
\usepackage{booktabs,siunitx,array,threeparttable}
\usepackage{upgreek}
\usepackage{orcidlink}

\usepackage[T1]{fontenc}
\usepackage{ae,aecompl}
\usepackage{appendix}

\usepackage{epsfig,amsfonts,mathrsfs,amsmath,amssymb,graphicx,color,slashed,multirow}
\usepackage{amsmath,latexsym,amssymb,graphicx,slashed,color,enumerate,url,cancel,gensymb}
\usepackage{textcomp}

\usepackage[paperwidth=210mm,paperheight=297mm,centering,hmargin=2.5cm,vmargin=2.5cm]{geometry}
\usepackage{tikz-feynman}
\tikzfeynmanset{compat=1.1.0}
\usetikzlibrary{patterns,decorations.pathreplacing}
\usetikzlibrary{decorations.pathmorphing}
\usetikzlibrary{decorations.markings}
\usetikzlibrary{arrows,shapes}
\usetikzlibrary{shapes.misc}
\tikzset{
  gluon/.style={decorate, draw=black,
    decoration={coil,amplitude=4pt, segment length=4pt,aspect=0.7}} 
}
\tikzset{
  photon/.style={decorate, decoration={snake}},
}

\usepackage{textgreek} 
\usepackage{booktabs} 

\hypersetup{colorlinks,citecolor= nicered,linkcolor= blue}
\definecolor{nicered}{rgb}{0.7,0.1,0.1}
\definecolor{nicegreen}{rgb}{0.1,0.5,0.1}

\definecolor{vdrgreen}{rgb}{0.0, 0.6, 0.0}

\definecolor{myblue}{cmyk}{0.65, 0.37, 0.0, 0.19}

\definecolor{blue(ncs)}{rgb}{0.0, 0.53, 0.74}

\AtBeginDocument{\hypersetup{citecolor=myblue,linkcolor=myblue,urlcolor=myblue}}

\begin{document}

\title{{Probing neutrino millicharges at the European Spallation Source}}
\author{Alexander Parada~\orcidlink{0000-0003-4864-7308}}\email{alexander.parada@esap.edu.co}
\affiliation{Escuela Superior de Administración Pública, Territorial Meta, Carrera 31A No.34A-23, C.P. 500001, Villavicencio, Colombia}
\author{G. Sanchez Garcia~\orcidlink{0000-0003-1830-2325}}\email{gsanchez@ific.uv.es}
\affiliation{Instituto de F\'{i}sica Corpuscular (CSIC-Universitat de Val\`{e}ncia), Parc Cient\'ific UV C/ Catedr\'atico Jos\'e Beltr\'an, 2 E-46980 Paterna (Valencia) - Spain}


\begin{abstract}
We study the potential of a set of future  detectors, proposed to be located at the  European Spallation Source (ESS), to probe neutrino millicharges through coherent elastic neutrino-nucleus scattering. In particular, we focus on detectors with similar characteristics as those that are under development for operation at the ESS, including detection technologies based on cesium iodine, germanium, and noble gases. Under the considered conditions, we show that the Ge detector, with a lighter nuclear target mass with respect to CsI and a noble gas like Xe, is more efficient to constrain neutrino millicharges, reaching a sensitivity of $\sim 10^{-9}e$ for diagonal neutrino millicharges, and $\sim 10^{-8}e$ for the transition ones. In addition, we study the effects of including electron scattering processes for the CsI detector, achieving an expected sensitivity of $\sim 10^{-10} e$ for the diagonal millicharges.   
\end{abstract}

\maketitle

\section{Introduction}
The observation of coherent elastic neutrino-nucleus scattering (CE$\upnu$NS) by the COHERENT collaboration in 2017~\cite{COHERENT:2017ipa} has raised the interest for the search of new physics in the neutrino sector at low energies. This process was originally proposed fifty years ago by Freedman ~\cite{PhysRevD.9.1389}, and so far has been confirmed by the COHERENT collaboration by using three different detection technologies, a CsI scintillating crystal \cite{COHERENT:2017ipa,COHERENT:2021xmm}, a single phase liquid argon (LAr) detector \cite{COHERENT:2020iec}, and more recently, a P-type point contact Ge detector \cite{Adamski:2024yqt}. In addition to the above detectors, the complete experimental program from the COHERENT collaboration includes a NaI scintillating crystal detector \cite{Akimov:2022oyb}. Furthermore, there are other  future experimental proposals that aim to use  neutrinos from spallation neutron sources to measure the process of CE$\upnu$NS from other different target materials. This is the case, for instance, of the proposed experiment at the European Spallation Source \cite{Abele:2022iml}, a facility that is currently under construction in Lund, Sweden, and that at its final stage will become the most powerful neutron beam source in the world. The proposed detection techniques at this facility include two similar targets as those from the COHERENT collaboration: a CsI cryogenic scintillator and a p-Type point contact Ge detector, but in addition they include a Xe high pressure gaseous TPC and an Ar scintillating bubble chamber \cite{Baxter:2019mcx}. In addition to the detection of CE$\upnu$NS using neutrinos from spallation sources, there are currently many experimental collaborations searching for signals of this interaction from other neutrino sources. For instance, the PandaX-4T \cite{PandaX:2024muv} and XENONnT~\cite{XENON:2024ijk} experiments have recently reported
the detection of CE$\upnu$NS from $^8$B solar neutrinos. Moreover, the Dresden-II reactor experiment \cite{Colaresi:2022obx} reported a CE$\upnu$NS signal that is strongly dependent on quenching factor measurements. Additionally,  many other collaborations are searching for CE$\upnu$NS signal from reactor neutrinos. To list some of them, we have: CONUS~\cite{Ackermann:2024kxo} (updated to CONUS$+$~\cite{magcevns:edgar}), CONNIE \cite{CONNIE:2021ggh, CONNIE:2024pwt}, $\upnu$GEN \cite{nGeN:2022uje}, Red-100 \cite{Akimov:2022xvr}, Ricochet \cite{Ricochet:2021rjo}, and NUCLEUS \cite{NUCLEUS:2022zti}. 

Since its first observation, a plethora of Standard Model (SM) properties and new physics scenarios have been tested by using the available CE$\upnu$NS data. For instance, a complete analysis combining the COHRERNT data from the CsI and LAr observations can be found in Refs.~\cite{DeRomeri:2022twg} and \cite{Cadeddu:2020lky}. The SM parameters that can be tested via CE$\upnu$NS include the weak mixing angle at low energies~\cite{AtzoriCorona:2023ktl} and the neutron root mean square (rms) radius of the target material~\cite{Cadeddu:2017etk}. Regarding new physics, the different scenarios for which CE$\upnu$NS is sensitive involve vector type Non-Standard Interactions (NSI) \cite{Denton:2020hop, Giunti:2019xpr, Miranda:2020tif}, and its interplay with the determination of the neutron rms radius \cite{canas2020interplay, Rossi:2023brv}, scalar and tensor type generalized neutrino interactions (GNI) \cite{AristizabalSierra:2018eqm, Lindner:2016wff, Flores:2021kzl}, neutrino electromagnetic properties \cite{Miranda:2019wdy, DeRomeri:2022twg, Cadeddu:2020lky}, scalar leptoquark scenarios \cite{Calabrese:2022mnp, DeRomeri:2023cjt} and even searches for dark fermions \cite{Candela:2023rvt}, among others. 

Neutrino electromagnetic properties include the widely studied effective neutrino magnetic moment \cite{Grimus:2000tq, AristizabalSierra:2021fuc}, neutrino charge radius \cite{Papavassiliou:2003rx}, and neutrino electric  millicharges (NEM) \cite{Foot:1989fh}, our topic of interest. The most restrictive limit on NEM, $q_{\nu}=3.0\times 10^{-21}e$, has been obtained by using a bound on the neutron charge and the neutrality of matter \cite{Raffelt:1999gv}. Regarding astrophysical measurements, a limit of $q_{\nu}=1.0\times 10^{-19}e$ was achieved based on the Neutrino Star Turning mechanism in supernova explosions \cite{Studenikin:2012vi}. Another of the most restrictive limits on NEM also comes from astrophysical measurements, $q_{\nu}=2.0\times 10^{-15}e$, by including analysis of neutrinos from SN 1987A \cite{Barbiellini:1987zz}. On the other hand, there has also been a significant effort to find constraints to this parameter from data of terrestrial neutrino experiments, where so far one of the most strong limit, $q_{\nu}=1.5\times 10^{-12}e$, was achieved from data of reactor neutrinos in the GEMMA experiment \cite{Studenikin:2013my}. Limits on NEM from combined analysis of experiments of elastic neutrino-electron interaction and future experimental proposals including CE$\upnu$NS were reported in Ref.~\cite{Parada:2019gvy}. Currently, CE$\upnu$NS bounds on NEM using spallation sources are weak \cite{DeRomeri:2022twg}. However, it has been shown that future CE$\upnu$NS experimental proposals using reactor sources can set limits on the NEM that can be competitive with neutrino-electron scattering bounds~\cite{Parada:2019gvy}. The rise of new neutrino detector technologies presents interesting new scenarios that will allow for sensitivity assessments on this parameter. Following the dynamics and expectations of new experimental proposals, in this paper we focus on the potential of the ESS to constrain neutrino millicharges, which complements the study of neutrino electromagnetic properties explored in Ref. \cite{Baxter:2019mcx}, where neutrino magnetic moments and neutrino charge radius were studied. 

The remaining of the paper is writen as follows. In section \ref{sec:framework} we give a brief overview of CE$\upnu$NS and neutrino millicharges, then in section \ref{sec:experiment} we describe the considered experimental setup, which is based on Ref. \cite{Baxter:2019mcx}, and the performed analysis in section \ref{sec:analysis}. Finally, we present our results and conclusions in sections \ref{sec:results} and \ref{sec:conclusions}, respectively.

\section{Theoretical Framework}\label{sec:framework}
In the following, we present the theoretical framework considered in this paper. We begin by introducing the process of CE$\upnu$NS and its main properties, and then we discuss the electromagnetic properties of neutrinos by mainly focusing on the millicharge extension to the SM.

\subsection{CE$\upnu$NS}
In a CE$\upnu$NS interaction, a neutrino interacts via neutral current with a nucleus as a whole, strongly enhancing the associated cross section with respect to other neutrino processes like, for instance,  neutrino-electron scattering and inverse beta decay. Within the SM framework, the CE$\upnu$NS cross section is flavor independent and it is given by~\cite{PhysRevD.9.1389}
\begin{equation}
\label{eq:cross}
\left (  \frac{\mathrm{d}\sigma}{\mathrm{d}T} \right )_{\textrm{SM}} = \frac{G_F^2M}{\pi}\left(1-\frac{MT}{2E_{\nu}^2}\right)\,F^2(q^2)\left(Q_{W}^V\right)^2,
\end{equation}
with $G_F$ the Fermi constant, $M$ the mass of the target material,  $E_\nu$ the energy of the incident neutrino, and $T$ the nuclear recoil energy. The quantity $Q_W^V$ is commonly called the weak charge,  and it is given by
\begin{equation}\label{eq:qweak}
    \left(Q_{W}^V\right)^2  = \left(Z\,g_V^p + N\,g_V^n \right)^2,
\end{equation}
where $g_V^p = 1/2 - 2\sin^2{\theta_W}$ and $g_V^n = -1/2$ are the coupling constants to protons and neutrons, respectively, defined within the SM, with $\theta_W$ the weak mixing angle. The enhancement of the CE$\upnu$NS cross section with respect to other processes can be seen from Eq. (\ref{eq:cross}). This is because $|g_V^p| << |g_V^n|$ and, consequently, the CE$\upnu$NS cross section effectively scales as $N^2$.

The function $F(q^2)$, with $q^2$ the momentum transfer, is called the nuclear form factor, which is introduced to describe the distribution of protons and neutrons within the nucleus. Among different parametrizations that can be found in the literature, here we use the Helm form factor~\footnote{We have explicitly checked that our results do not significantly change by the election of the form factor.}, given by \cite{Helm:1956zz}
\begin{equation}
F(q^2) = 3\, \frac{j_1(qR_0)}{qR_0}\,e^{-q^2s^2/2},
\end{equation}
where $j_1$ stands for the spherical Bessel function of order one, $s$ is the surface thickness, which we take as 0.9 fm \cite{Lewin:1995rx}. The parameter $R_0$ in the above equation is related to the neutron rms radius\footnote{The values used for $R$ in our analysis are given in Table \ref{tabless}.}, $R$, through $R_0^2=5(R^2-3s^2)/3$~\cite{Sierra:2023pnf}. This dependency has been exploited to study the neutron rms radius from CE$\upnu$NS interactions \cite{Cadeddu:2017etk}.

\subsection{Neutrino-Electron scattering}
As we have discussed, our main purpose is the study of the CE$\upnu$NS process at the ESS. However, we will see that for new physics scenarios as neutrino millicharges, it might be important to consider the effects of electron scattering at low energy thresholds. For an electron on an atomic nucleus $\mathcal{A}$, the neutrino-electron scattering cross section is flavor dependent, and it is given by~\cite{Vogel:1989iv},
\begin{equation}
\left (  \frac{\mathrm{d}\sigma}{\mathrm{d}T} \right )_{\textrm{SM}} = Z_{\textrm{eff}}^\mathcal{A}(T) \frac{G_F^2m_e}{2\pi}\left[\left ( g_V^{\nu_\alpha} + g_A^{\nu_\alpha} \right )^2 + \left ( g_V^{\nu_\alpha} - g_A^{\nu_\alpha} \right )^2\left ( 1 - \frac{T}{E_\nu} \right )^2  - \left ( \left (  g_V^{\nu_\alpha}\right )^2 - \left ( g_A^{\nu_\alpha}\right )^2 \right )\frac{m_eT}{E_{\nu}^2}\right]
\label{eq:cross:es}
\end{equation}
where $m_e$ is the mass of the electron, and $g_{V}^{\nu_\alpha}$ and $g_{A}^{\nu_\alpha}$ are the flavor-dependent coupling constants defined within the SM. For an incoming electron neutrino, $\nu_e$, we have $g_{V}^{\nu_e} = 2\sin^2(\theta_W) + 1/2$ and $g_{A}^{\nu_e} = 1/2$, while for an incoming muon neutrino we have $g_{V}^{\nu_\mu} = 2\sin^2(\theta_W) - 1/2$ and $g_{A}^{\nu_\mu} = -1/2$. The difference between the two cases stems from the fact that, for an incoming electron neutrino, the cross section has contributions from both charged and neutral currents, while in the case of a muon neutrino we only have neutral current contributions. Additionally, when considering an incoming antineutrino there is a change of sign in the axial contribution, which means that for the muon antineutrinos at the ESS we have $g_{V}^{\overline{\nu}_\mu} = 2\sin^2(\theta_W) - 1/2$ and $g_{A}^{\overline{\nu}_\mu} = +1/2$. Finally, note that Eq.~(\ref{eq:cross:es}) depends on an effective number of protons, $Z_{\textrm{eff}}^\mathcal{A}$, seen by the incoming neutrino as a function of the deposited energy $T$. The detailed shape of this effective number of protons can be found in Ref. \cite{DeRomeri:2022twg}.

\subsection{Neutrino Millicharges}

Within the SM, neutrinos are  massless and chargeless particles. However, the discovery of neutrino oscillations has provided an unambiguous proof that neutrinos indeed have mass, being now necessary to extend the SM to explain them. In some of these extensions, neutrinos acquire electromagnetic properties at a loop level, being of particular interest since they can, for instance, distinguish between Dirac or Majorana neutrinos \cite{Broggini:2012df}. Among the electromagnetic properties that neutrinos can acquire in theories beyond the SM, we have magnetic and electric dipole moments, whose values can be sensitive to neutrino mass, electric millicharges, and charge radius, although the latter is also present in SM neutrinos. To introduce these properties, we begin by considering the interaction of a neutrino with a photon, which amplitude can be represented by the matrix element~\cite{Giunti:2014ixa}
\begin{equation}
  \langle \nu(p_{f},s_{f})|J^{EM}_{\mu}|\nu(p_{i},s_{i})\rangle = i\bar{u}_{f}\Gamma_{\mu}(q)u_{i},
\label{Eq:1}  
\end{equation}
where $p_{i}, p_{f}$ represent the four-moment and $s_{i}, s_{f}$ the spin projections of the initial and final neutrino states, respectively \cite{Kayser:1982br,Giunti:2014ixa}.
In its most general form, consistent with Lorentz and SM gauge symmetries, the vertex function, $\Gamma_\mu$, in Eq. (\ref{Eq:1}) is represented in terms of four form factors~\cite{Giunti:2014ixa}
\begin{eqnarray}
  \Gamma_{\mu}(q) = F_{Q}(q^{2})\gamma_{\mu} + F_{A}(q^{2})(q^{2}\gamma_{\mu}-2miq_{\mu})\gamma_{5}\\
  \nonumber + F_{M}(q^{2})\sigma_{\mu\nu}q_{\nu} + F_{E}(q^{2})i\sigma_{\mu\nu}q_{\nu}\gamma_{5},
\label{Eq:2}  
\end{eqnarray}  

\noindent where $F_{Q}(q^{2})$, $F_{A}(q^{2})$, $F_{M}(q^{2})$, and $F_{E}(q^{2})$ are $4\times4$ matrices called real charge, anapole, magnetic dipole, and electric dipole moment form factors, respectively. For zero momentum transfer (real photons), the form factors above can be interpreted as the neutrino millicharge, anapole
moment, magnetic moment, and electric dipole moment. For definiteness, we denote these quantities, respectivley, as
\begin{equation}
  F_{Q}(0) \equiv q_{\nu},~~~\;F_{A}(0) \equiv a_\nu,~~~\;F_{M}(0) \equiv \mu_{\nu},\;~~~F_{E}(0)\equiv d_\nu.
\label{Eq:3}  
\end{equation}

Here we focus our attention on the neutrino millicharge, $q_{\nu}$. For a complete discussion of the physical meaning of each of the terms defined in Eq.~(\ref{Eq:3}), we refer the reader to Ref.~\cite{Giunti:2014ixa}. To motivate the concept of a neutrino millicharge, lets recall that within the standard theory of SU$(2)_{L}\times$ U$(1)_{Y}$ electroweak interactions, the electric charge, $Q$, of a particle is related to the weak isospin, $I_{3}$, and the hypercharge, $Y$,  by the Gell-Mann–Nishijima relation, $Q~=~I_{3}~+~Y/2$. For the gauge theory to be consistent, it must be free of triangle anomalies so that renormalizability is guaranteed. Then, triangle diagrams must be cancelled among the fermionic content present in the theory. Within the SM, this cancellation is given between left handed fields and their right handed counterparts. However, in this theory neutrinos are treated only as left-handed particles (with isospin $I_{3}=1/2$), and anomaly cancellation requires $Y=-1$, implying charge quantization and, from the above relation, that for neutrinos $Q=0$. 

This simple argument for the neutrality of neutrinos applies when considering only left handed neutrinos. However, neutrinos are massive, and many theories intended to explain their masses need to introduce the neutrino right handed counterparts. In such case, anomaly cancellation is given independently of $Y$, and now the neutrino hypercharge is no longer fixed\footnote{The reader can find a more detailed description about theoretical basis for the neutrino millicharge in \cite{Fukugita:2003en}, \cite{Giunti:2014ixa} and references therein.}. This leads to a significant consequence regarding electric charge quantization, which can be slightly altered, implying a small charge, $Q = \varepsilon$, for neutrinos. The new non standard hypercharge can be associated to a new anomaly free U$(1)$ symmetry like, for instance $U(1)_{B-L}$, which indeed becomes anomaly free by introducing a right-handed neutrino, $\nu_{R}$, in a minimally extended scenario. 


\subsection{Effects of neutrino millicharges on neutrino interactions}
Allowing for a neutrino millicharge has an impact on the computation of the CE$\upnu$NS cross section.
These effects have been widely studied and depend on whether we consider diagonal electric charges, $q_{\alpha\alpha}$, or transition electric charges, $q_{\alpha\beta}$, with $\alpha\neq\beta$. For the case of diagonal terms, the neutrino electric charge effects  may be interpreted as a redefinition of the coupling $g_V^p$ present in the SM cross section of CE$\upnu$NS, such that \cite{Kouzakov:2017hbc}
\begin{equation}
    g_V^p \rightarrow g_V^p + \frac{\sqrt{2}\pi\alpha_{\textrm{EM}}}{G_F MT}q_{\nu_{\alpha\alpha}},
\end{equation}
 with $\alpha_{\textrm{EM}}$ the fine structure constant. This redefinition means that there is an interference between SM and electric charge contributions. In contrast, for transition electric charges there is not such interference and the cross section contributions add up independently.
All together, the weak charge introduced in Eq. \eqref{eq:qweak} can be redefined such that, in the presence of a neutrino millicharges, we have \footnote{Even though neutrinos and antineutrinos have opposite charges, the shift in the coupling constant has the same sign for neutrinos and antineutrinos.}
\begin{equation}\label{eq:qmilli}
\left(Q_{\textrm{MC}}^V\right)^2  = \left( Q_{\textrm{W}}^V\ + \frac{\sqrt{2}\pi\alpha_{\textrm{EM}} Z}{G_F M} \frac{q_{\nu_{\alpha\alpha}}}{T} \right)^2 + \left( \frac{\sqrt{2}\pi\alpha_{\textrm{EM}} Z}{G_F M} \frac{|q_{\nu_{\alpha\beta}}|}{T} \right)^2.
\end{equation}
By expanding this modified weak charge, we can see that when millicharges are present, there are three contributions to the CE$\upnu$NS cross section: the usual  SM term, an interference term between the SM and the millicharge contribution, which scales as $q_\nu/T$, and a purely electromagnetic term that goes as $q_\nu^2/T^2$. As we will discuss later, the dependence on powers of $q_\nu/T$ of the interference and purely electromagnetic terms result on an enhancement of the cross-section at low energy thresholds. This will impact the sensitivity of our detectors to the millicharge under study. Similarly, for the case of neutrino-electron scattering, the impact of the diagonal millicharge contribution is a redefinition of the coupling constant $g_V^{\nu_\alpha}$, which is now given by
\begin{equation}
    g_V^{\nu_\alpha} \rightarrow g_V^{\nu_\alpha} - \frac{\sqrt{2}\pi\alpha}{G_F m_eT}q_{\nu_{\alpha\beta}}.
\end{equation}
Again, the impact of this change is the presence of an interference term between the SM and millicharge contribution, which also goes as $q_\nu/T$, and a purely electromagnetic contribution to the cross section proportional to 
$q_\nu^{2}/T^{2}$, in a similar way to the CE$\upnu$NS but with different constant factors in each term \cite{Giunti:2014ixa}.

\section{Experimental setup}\label{sec:experiment}


\subsection{European Spallation Source}
At its full power, the ESS will provide the most intense neutron beams in the world \cite{Baxter:2019mcx}, which are produced as a result of the collision of a high energy proton beam with a tungsten target. Neutrinos are generated as a byproduct of these interactions and can be separated in two different sets regarding their production time. On the one hand, we have "prompt" neutrinos, which are monoenergetic muon neutrinos that result from the decay of positively charged pions at rest. In this case, the neutrino flux can be analitically described by a delta function
\begin{equation} 
\frac{d N_{\nu_\mu}}{d E_\nu}(E_\nu)  = \eta \, \delta\left(E_\nu-\frac{m_{\pi}^{2}-m_{\mu}^{2}}{2 m_{\pi}}\right),\,
\label{eq:prompt:flux}
\end{equation}
where $\eta$ is a normalization constant (see below), $E_\nu$ is the neutrino energy, and $m_\pi$ and $m_\mu$ are the pion's and muon's mass, respectively. In addition, within the target material we have the production of charged  muons, which eventually decay also at rest, producing muon antineutrinos and electron neutrinos. The distributions of these particles are widely known and they are given, respectively, by
\begin{equation}
\frac{d N_{\bar{\nu}_\mu}}{d E_\nu}(E_\nu) = \eta \frac{64 E^{2}_\nu}{m_{\mu}^{3}}\left(\frac{3}{4}-\frac{E_\nu}{m_{\mu}}\right),
\label{eq:delayed:1}
\end{equation}
\begin{equation}
\frac{d N_{\nu_e}}{d E_\nu}(E_\nu)  = \eta \frac{192 E^{2}_\nu}{m_{\mu}^{3}}\left(\frac{1}{2}-\frac{E_\nu}{m_{\mu}}\right)  \, .
\label{eq:delayed:2}
\end{equation}
The normalization factor in the above expressions depends on the spallation source used for the experiment and, in general, is given by $\eta = rN_{\textrm{POT}}/4\pi L^2$, with $r$ the number of produced neutrinos per flavour, $N_{\textrm{POT}}$ the number of protons on target, and $L$ the distance from the production point to the detector. At its full capacity, the ESS is expected to operate with a beam power of $5$ MW and an energy of 2 GeV, resulting in a total $N_{\textrm{POT}}$ of $2.8\times10^{23}$ per year~\cite{Baxter:2019mcx}, where we have considered a total of 5000 operation hours per calendar year. Under this condition, a value of $r = 0.3$ is expected \cite{Baxter:2019mcx}, and we use it for all of our computations.
Once the neutrino source is known, then a detector is needed to compare an expected number of events with the experimental result. For the particular case of the ESS, an initial experimental proposal for CE$\upnu$NS measurements considered six different detection materials, including cesium iodine (CsI), germanium (Ge), xenon (Xe), argon (Ar), silicon (Si), and octafluorpropane (C$_3$F$_{8}$) detectors~\cite{Baxter:2019mcx}. Moreover, it has recently been announced that the CsI and Ge detectors, together with a noble gas TPC, are in active development for deployment in the upcoming years \cite{Simon:2024xwb, Monrabal_M7_2024}. For our analysis, we will consider these three detectors, with Xe as the noble gas, taking their main characteristics from Ref. \cite{Baxter:2019mcx}, and summarized in Table \ref{tabless}, where, among other relevant quantities, we show the detectors mass and expected thresholds. In all cases, we assume our detectors to be located at a distance of 20 m from the source. 

\begin{table}[b]
\centering
\begin{tabular}{|c|c|c|c|c|c|} 
\hline
\textbf{~~Target~~} & \textbf{~~$R$ (fm)~~} & \textbf{~~Mass (kg)~~} & \textbf{~~Threshold (keV)~~} & \textbf{~~Background (ckkd)~~} & \textbf{~~$\sigma_0$~~}  \\ 
\hline
CsI            & 4.83\footnote{The rms radius is the same for Cs and I. }            & 22.5               & 1.0                      &  10                         & 0.3             \\ 
\hline
Xe          & 4.79               & 20                 & 0.9                      &  10                         & 0.36              \\ 
\hline
Ge            & 4.06             & 7                  & 0.6                      &  3\footnote{The Ge background in Ref. \cite{Baxter:2019mcx} is given in keV$_{ee}$, which we convert to keV$_{nr}$ by using a quenching factor of 0.2 given in the same reference.}                        & 0.09              \\
\hline
\end{tabular}
\caption{ Main characteristics of the detectors considered for this work  \cite{Baxter:2019mcx}.}
\label{tabless}
\end{table}

\section{Analysis}\label{sec:analysis}

In general, the predicted number of events for a neutrino counting experiment is computed by the convolution of the differential cross section with the neutrino flux. For a spallation source as the ESS, the total neutrino flux, $dN_\nu/dE_\nu$, is given by the sum of the individual fluxes given in Eqs.~(\ref{eq:prompt:flux}) to (\ref{eq:delayed:2}). Then, the expected event rate, for the $i$th bin, in our considered detectors, is given by
\begin{align}\label{eq:Nevents_CEvNS_alpha}
N_{i}^{\mathrm{CE}\nu\mathrm{NS}}
= 
\mathcal{N}
\int_{T_{i}}^{{T}_{i+1}}
\hspace{-0.3cm}
d T\,
\epsilon(T)
\int_{0}^{T^{\prime\text{max}}}
\hspace{-0.3cm}
dT'
\,
G(T,T')  \int_{E_\nu^{\text{min}}(T')}^{E_\nu^{\text{max}}}
\hspace{-0.3cm}
d E_\nu
\frac{d N_{\nu}}{d E_\nu}(E_\nu)
\frac{d\sigma}{dT'}(E_\nu, T'),
\end{align}
where $T$  is the reconstructed recoil energy and $T'$ is the real recoil energy of the nucleus. In the previous equation, the lower limit of the first integral is determined by kinematic constraints, and it is given by $E_{\nu}^{min} = \sqrt{MT'/2}$, while the upper limit, $E_\nu^{max}$, is given by the maximum neutrino energy from the source, which is around 52.8 MeV. We have also taken into account the resolution of the experiment through a gaussian smearing function $G(T,T')$, for which, by following the procedure in Ref. \cite{Baxter:2019mcx}, we consider an energy-dependent width given by $\sigma = \sigma_0 \sqrt{TT_{\textrm{Th}}}$, with $\sigma_0$ and $T_{\textrm{Th}}$ given in Table \ref{tabless}. Additionally, we have taken into account a constant detector efficiency, $\epsilon(T)$, of 80\% as in Ref. \cite{Baxter:2019mcx}. Under these assumptions, we obtain the same  SM event distributions as in Ref.~\cite{Chatterjee:2022mmu}. To explore the sensitivity of our considered experimental setup to neutrino millicharges, we  minimize the least squares function
\begin{equation}
  \chi^2 = 2\sum_{i}\left [ N^{\textrm{th}}_{i}(q_{\nu_{\alpha\beta}}) - N^{\textrm{exp}}_{i} + N^{\textrm{exp}}_{i}\ln\left ( \frac{N^{\textrm{exp}}_{i}}{N^{\textrm{th}}_{i}(q_{\nu_{\alpha\beta}})} \right )\right ] + \frac{\alpha_0}{\sigma_{\alpha_0}^2} + \frac{\beta_0}{\sigma_{\beta_0}^2},
    \label{eq:chi:future}
\end{equation}
where we have defined
\begin{equation}
    N^{\textrm{th}}_i(q_{\nu_{\alpha\alpha}}) = (1+\alpha_0) N^{\textrm{CE$\nu$NS}}_i(q_{\nu_{\alpha\alpha}}) +  (1+\beta_0)N^{\textrm{bckg}}.
    \label{N:chi:future}
\end{equation}
Here the $i$ index runs over the recoil energy bins on which our data is divided, which is taken from Ref.~\cite{Chatterjee:2022mmu} for each detector. In Eq. \eqref{eq:chi:future}, the quantity $N_i^{\textrm{th}}$ represents the predicted number of events when considering a non-zero value of the neutrino millicharge, $q_{\nu_{\alpha\beta}}$. This theoretical prediction is given by the sum of CE$\upnu$NS events, denoted as $N_i^{\textrm{CEvNS}}$, and expected backgrounds, denoted as $N_i^{\textrm{bckg}}$. For the latter, we consider the background model used in Ref. \cite{Baxter:2019mcx}, which is given in terms of counts per kg per keV per day (ckkd) as shown in Table \ref{tabless} for each detector. Coming back to Eq.~\eqref{eq:chi:future}, $N_i^{\textrm{exp}}$ represents the expected experimental measurement, which we will consider the SM prediction since we are studying the sensitivity of a future experimental array. 
The $\chi^2$ function is minimized over the nuisance parameters $\alpha_0$ and $\beta_0$, which are associated to CE$\upnu$NS and background signals, respectively. Each of these parameters has an associated uncertainty, with values $\sigma_{\alpha_0} = 10\%$ and $\sigma_{\beta_0} = 1\%$, as taken from Ref. \cite{Baxter:2019mcx},

\section{Results}\label{sec:results}
In this section we present our main results for the expected sensitivity of the ESS to constrain neutrino millicharges.

\subsection{Single neutrino millicharge }\label{sec:results:CEvNS}

\begin{figure}[t]
\centering
\includegraphics[width=0.49\textwidth]{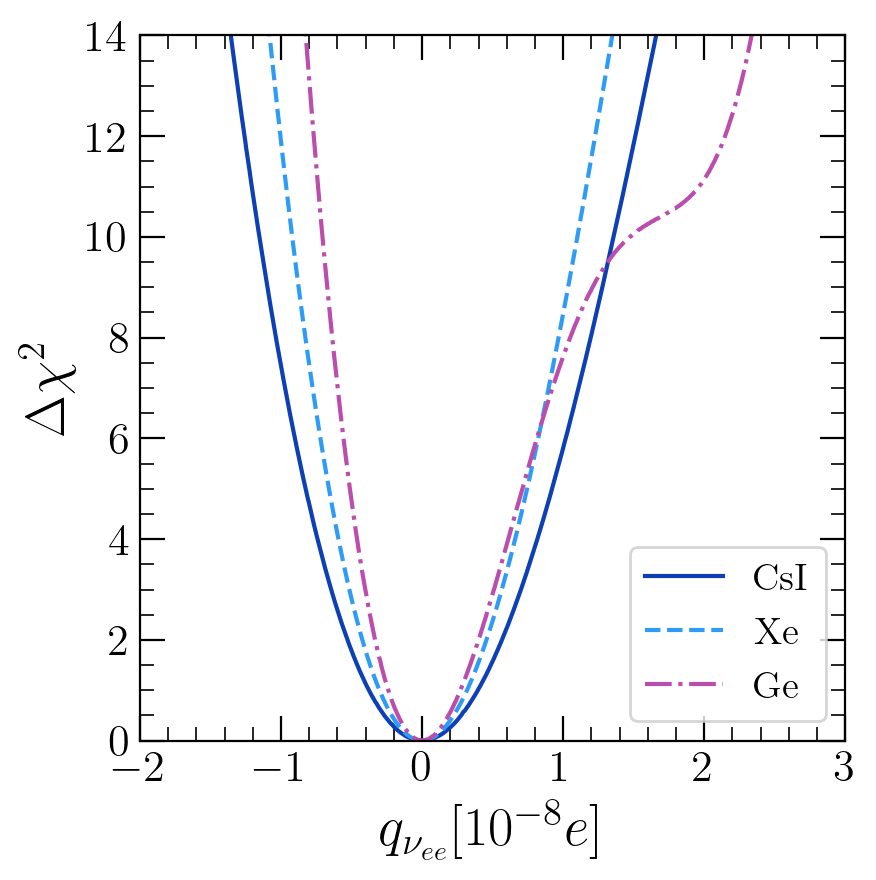}
\hspace{0.1cm}
\includegraphics[width=0.49\textwidth]{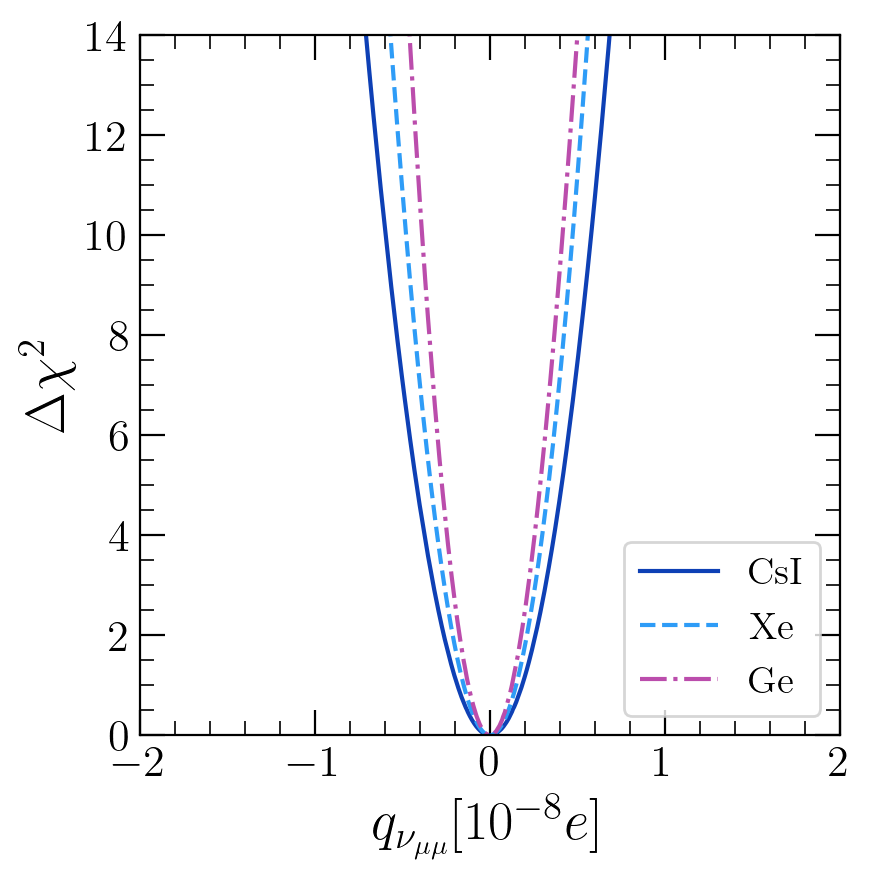}
\caption{Expected one dimensional $\Delta\chi^2$ profile for $q_{\nu_{ee}}$ (left panel) and $q_{\nu_{\mu\mu}}$ (right panel). We assume three years of data taking for three different detectors located at the ESS: CsI (solid blue), Xe (dashed cyan), and Ge (dash-dotted violet). }
\label{fig:1D:results}
\end{figure}

We begin our discussion with the results when allowing only one $q_{\nu}$ parameter to be  different from zero at a time. For now we focus on diagonal neutrino charges and we show the expected sensitivity to  $q_{\nu_{ee}}$ in the left panel of Fig.~\ref{fig:1D:results}, where we distinguish the results for CsI (solid blue), Xe (dashed cyan), and Ge (dash-dotted violet). The corresponding expected constraints at a 90\% C.L. are given in Table \ref{tab:qee}, with the most restrictive expected bound obtained from the Ge detector, constraining the millicharge to be, in units of the fundamental charge $e$, within  the interval $[-3.9, 4.9]\times 10^{-9}e$. From this result, notice that the detectors at the ESS would be expected to lower approximately two orders of magnitude the current constraint obtained from the combination of CsI and LAr detectors from COHERENT, which from the analysis presented in Ref.~\cite{DeRomeri:2022twg} is $q_{\nu_{ee}} \in [-1.0, 1.3] \times 10^{-7}e$. A similar conclusion is obtained for the case of $q_{\nu_{\mu\mu}}$ from the right panel of Fig.~\ref{fig:1D:results}, where we use the same color code as in the previous analysis. Again, the corresponding expected sensitivities are given in Table \ref{tab:qee}, with the most restrictive constraint expected from the Ge detector, with $q_{\nu_{\mu\mu}} \in [-2.1, 2.1] \times 10^{-9}e$. We can compare with the current constraint of $q_{\nu_{\mu\mu}} \in [-5.1, 4.6] \times 10^{-8}e$ as obtained in Ref. \cite{DeRomeri:2022twg} from the combined COHERENT data, concluding that the ESS is expected to lower this bound by one order of magnitude.

\begin{table}[b]
\centering
\begin{tabular}{|c|c|c|c|c|c|}
\hline
~~Detector~~ & ~~$q_{\nu_{ee}} [\times 10^{-9}e]$~~ & ~~$q_{\nu_{\mu\mu}} [\times 10^{-9}e]$~~ & ~~$q_{\nu_{e\mu}} [\times 10^{-8}e]$~~ & ~~$q_{\nu_{\mu\tau}} [\times 10^{-8}e]$~~ & ~~$q_{\nu_{e\tau}} [\times 10^{-8}e]$~~  \\ 
\hline
CsI            & $[-6.1, 6.6]$                 & [-3.0, 3.0] & [-1.2, 1.2] & [-1.5, 1.5]   & [-2.2, 2.2] \\
\hline
Xe            & [-4.9, 5.3]                 & [-2.5, 2.5]  & [-1.0, 1.0] & [-1.2, 1.2]    & [-1.7, 1.7]     \\
\hline
Ge                  & [-3.9, 4.9]                 & [-2.1, 2.1] & [-0.6, 0.6] & [-0.7, 0.7] & [-1.0, 1.0] \\
\hline
\end{tabular}
\caption{Expected sensitivities at 90\% C.L. to neutrino millicharges at the ESS.}
\label{tab:qee}
\end{table}

\begin{figure}[t]
\centering
\includegraphics[width=0.49\textwidth]{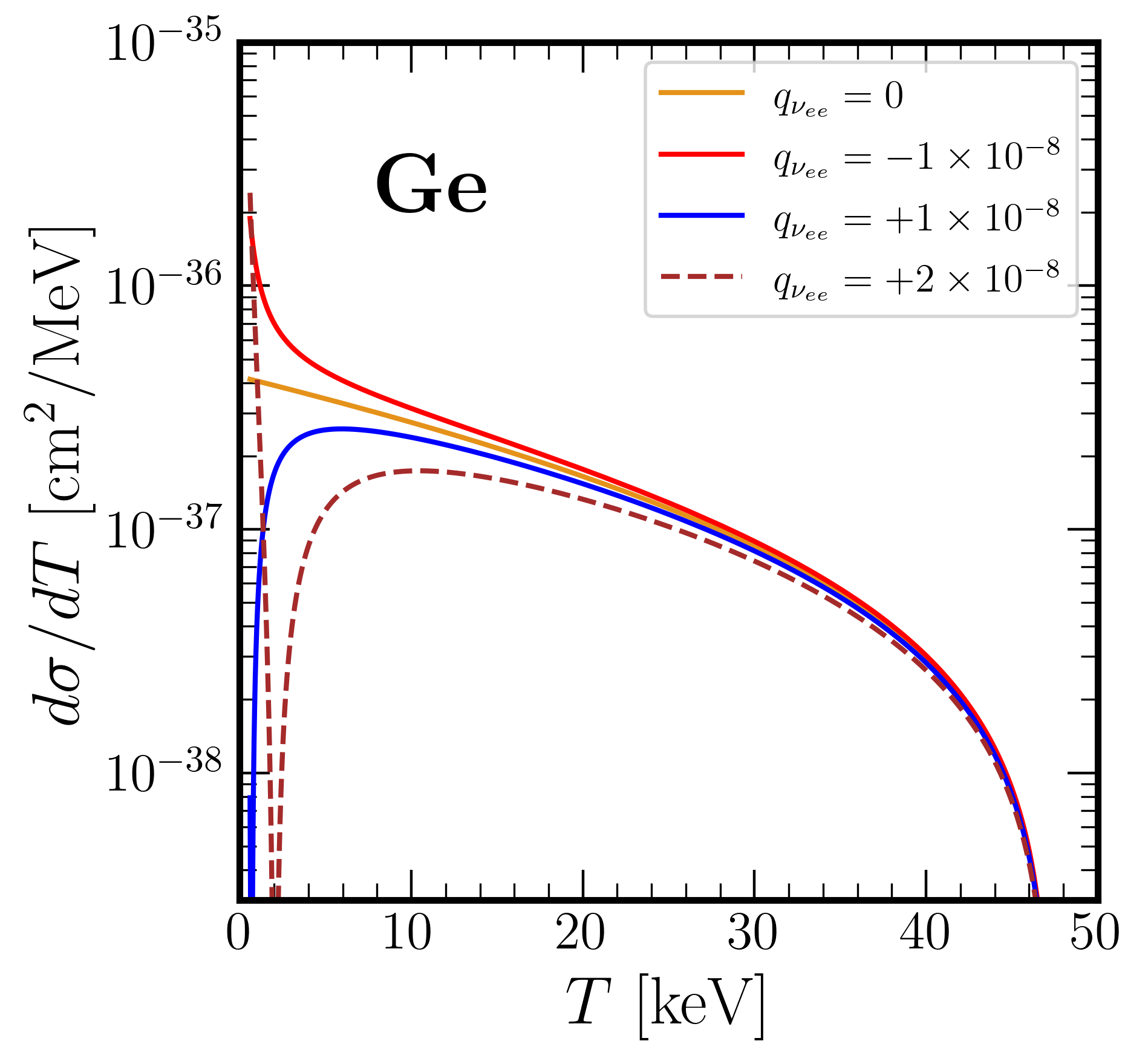}
\hspace{0.1cm}
\includegraphics[width=0.49\textwidth]{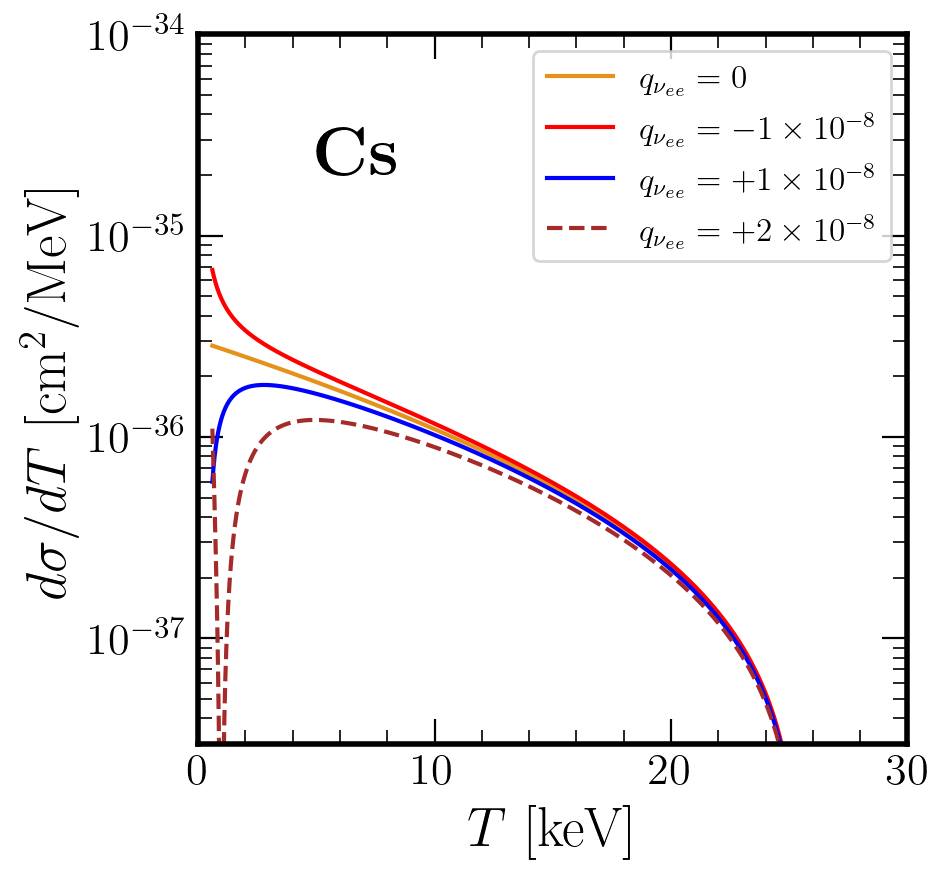}
\caption{Predicted cross section, as a function of $T$, for Ge (left panel) and Cs (right panel) targets. We compare the SM result (orange) with the result obtained by considering different millicharge values: $q_{\nu_{ee}} = -1\times 10^{-8}$ in red, $q_{\nu_{ee}} = +1\times 10^{-8}$ in blue, and $q_{\nu_{ee}} = +2\times 10^{-8}$ in dashed brown. In all cases we assume a fixed incoming neutrino energy of 40 MeV.}
\label{fig:cross:results}
\end{figure}

There are two features from our results that are important to remark. First, the qualitative behaviour of the $\Delta\chi^2$ profile for Ge seen in the left panel of Fig. \ref{fig:1D:results}, which is different from the other two in this range, and second, the fact that among the three considered detectors, it is the Ge one that gives the better constraints. This may seem  counter intuitive since Ge is the detector with less mass, and hence, from which we have less statistics. However, both of these features have a similar origin and can be explained from the two panels in Fig. \ref{fig:cross:results}, where we show the  CE$\upnu$NS cross section as a function of the nuclear recoil energy, $T$, for a fixed $E_\nu = 40$~MeV, and for two different target materials, Ge in the left panel and Cs in the right panel\footnote{The cross sections for I and Xe have the same behaviour as for Cs.}. In both cases, we compare the SM cross section (solid orange) with the modified one that results when considering non-zero millicharges, taking three different values of $q_{\nu_{ee}}$.

From the weak charge defined in Eq.~(\ref{eq:qweak}), we notice that, since $|g_V^p| << |g_V^n|$, within the SM, the value of $Q_W^V$ is negative. When all the other charges are set to zero, the effect of a negative $q_{\nu_{ee}}$ on the modified cross section in Eq.~(\ref{eq:qmilli}) is that $Q_{\textrm{MC}}^V$  will become more negative when compared to the SM case, enhancing the CE$\upnu$NS cross section, specially at low values of $T$. This is illustrated in Fig. \ref{fig:cross:results}, where the red lines in the left (Ge) and right (Cs) panels correspond to $q_{\nu_{ee}} = -1\times10^{-8}e$. We see that now the cross section is above the SM prediction for every $T$, increasing the expected number of events with respect to the SM prediction. 
The more negative the value of $q_{\nu_{ee}}$, the more predicted events, and hence the more we deviate from the SM prediction, increasing  the value of $\Delta\chi^2$, according to Eq.~\eqref{eq:chi:future}. 
Turning now to a positive millicharge, and focusing on the Ge detector, we consider the case on which $q_{\nu_{ee}}$ is greater than zero, but relatively small. Then, the second term within parenthesis in Eq.~(\ref{eq:qmilli}) is positive, but lower in absolute value than $Q_W^V$ for all $T$. Since this is true for the whole region of interest, the predicted cross section is overall lower than the SM prediction. This can be seen from the blue lines in the left panel of Fig. \ref{fig:cross:results}, which  correspond to $q_{\nu_{ee}} = +1\times10^{-8}e$ and fall below the SM cross section (orange lines). 
Approximately up to this millicharge value, the larger the millicharge, the lower cross section, and the less events we predict. Then, the $\Delta\chi^2$ gradually increases in the range $[0, 1\times10^{-8}]e$, as we see in the left panel of Fig. \ref{fig:1D:results}. 

Keeping the discussion on the Ge detector, we now consider intermediate positive values of the millicharge, for instance, $q_{\nu_{ee}} = +2.0\times10^{-8}e$, illustrated as dashed lines in Fig. \ref{fig:cross:results}. For very small $T$, the second term in Eq.~(\ref{eq:qmilli}) will be positive and large  enough so that $Q_{\textrm{MC}}^V > |Q_W^V|$, increasing the predicted cross section with respect to the SM. However, as $T$ increases, the second term in Eq.~(\ref{eq:qmilli}) decreases, getting to a critical value, $T_{c}$, where the same term cancels that of $Q_W^V$, giving as a result that $Q_{\textrm{MC}}^V = 0$. From the left panel in Fig. \ref{fig:cross:results}, we see that this critical value is of $T_c = 2$ keV for an incoming neutrino energy of 40 MeV on a Ge target. As $T$ keeps increasing, the second term in Eq. (\ref{eq:qmilli}) goes to zero and we recover again the SM cross section, as seen from above~$\approx$ 40 keV for Ge in the same figure, where the dashed line overlaps with the SM prediction (orange line). We conclude that for large enough positive values of $q_{\nu_{ee}}$, there is a small interval on $T$ where more events than the SM are predicted, while there is another where we predict less, restricting the $\Delta\chi^2$ from rapidly increasing. This explains the behaviour of the $\Delta\chi ^2$ profile for Ge in the range $[1,2]\times10^{-8}e$. In our example, when $q_{\nu_{ee}} > 2\times 10^{-8}e$, the cross section is now above the SM prediction, covering a larger interval on $T$, increasing the expected number of events with respect to the SM prediction in this interval. Hence, the $\Delta\chi^2$ profile increases at a faster rate for $q_{\nu_{ee}} > 2\times 10^{-8}e$, as shown in the left panel of Fig. \ref{fig:1D:results}. 

Since CsI and Xe are heavier nuclei, the above discussed positive values of $q_{\nu_{ee}}$  are such that the cross section is still less than the SM prediction for all $T$. This can be seen on the right panel of Fig. \ref{fig:cross:results}. However, it is worth mentioning that the same effect for the $\Delta\chi^2$ profile as for the Ge case can be found but at larger millicharges and it is not displayed in the figure. The same situation is found for the three analyzed detectors for large values of $q_{\nu_{\mu\mu}}$, which are not displayed in the figure.
Once we have discussed the behaviour of the $\Delta\chi ^2$ profiles, it is also important to understand why the Ge detector gives the best sensitivity to millicharges. As we mentioned before, one would expect the CsI and Xe detectors to give better constraints since they are larger in both detector mass and neutron number, having a larger statistics for the SM cross section\footnote{In fact, for our experimental setup, we expect around ten times more statistics from the Xe and CsI detectors than from the Ge one, as can be seen from the distributions in Fig. 1 of Ref.~\cite{Chatterjee:2022mmu} }. However, there is another important aspect to consider. First, notice that the millicharge contribution to the cross section becomes relevant at low nuclear recoil energies. Then, coming back to Fig. \ref{fig:cross:results}, we compare, for instance, the orange lines (SM prediction) with the red lines ($q_{\nu_{ee}} = -1\times10^{-8}e$) on each panel. We notice that at, lets say, $T = 1$ keV, the cross section for $q_{\nu_{ee}} = -1\times10^{-8}e$ changes by a factor of 2 with respect to the SM for the case of Cs, while for the Ge detector it changes about a factor of 5. This means that for the same fixed value of the millicharge, the proportion that we deviate from the SM prediction is more for Ge than for CsI, resulting in a larger value of $\Delta\chi^2$ (see Eq. \eqref{eq:chi:future}) for the Ge detector, and hence, a better constraint. In addition, as shown in Table \ref{tabless}, this detector is expected to have a lower threshold (0.6 keV), than either the CsI (1 keV) or the Xe (0.9 keV), enhancing more the contribution of the millicharge cross section for Ge. Hence, we conclude that the proposed Ge detector is expected to give the best sensitivity when trying to constrain neutrino millicharges. This is opposed to the case, for instance, of NSI, where large detectors, with heavy nuclei give better constraints~\cite{Chatterjee:2022mmu}.

\begin{figure}[t]
\centering
\includegraphics[width=0.31\textwidth]{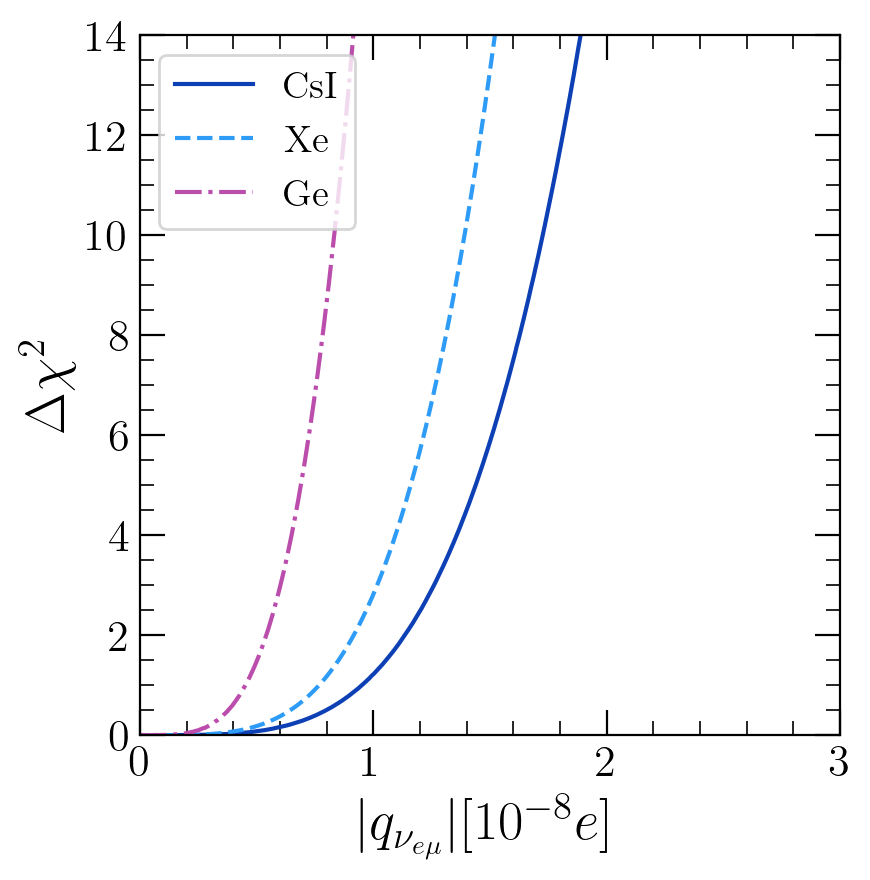}
\hspace{0.1cm}
\includegraphics[width=0.31\textwidth]{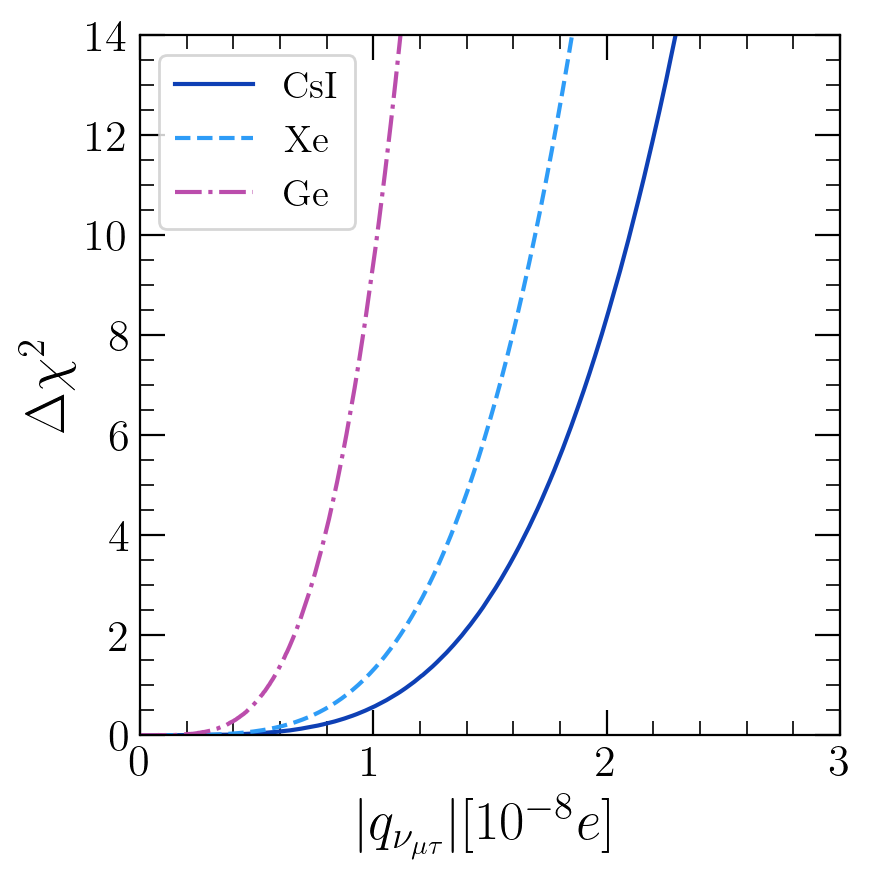}
\hspace{0.1cm}
\includegraphics[width=0.31\textwidth]{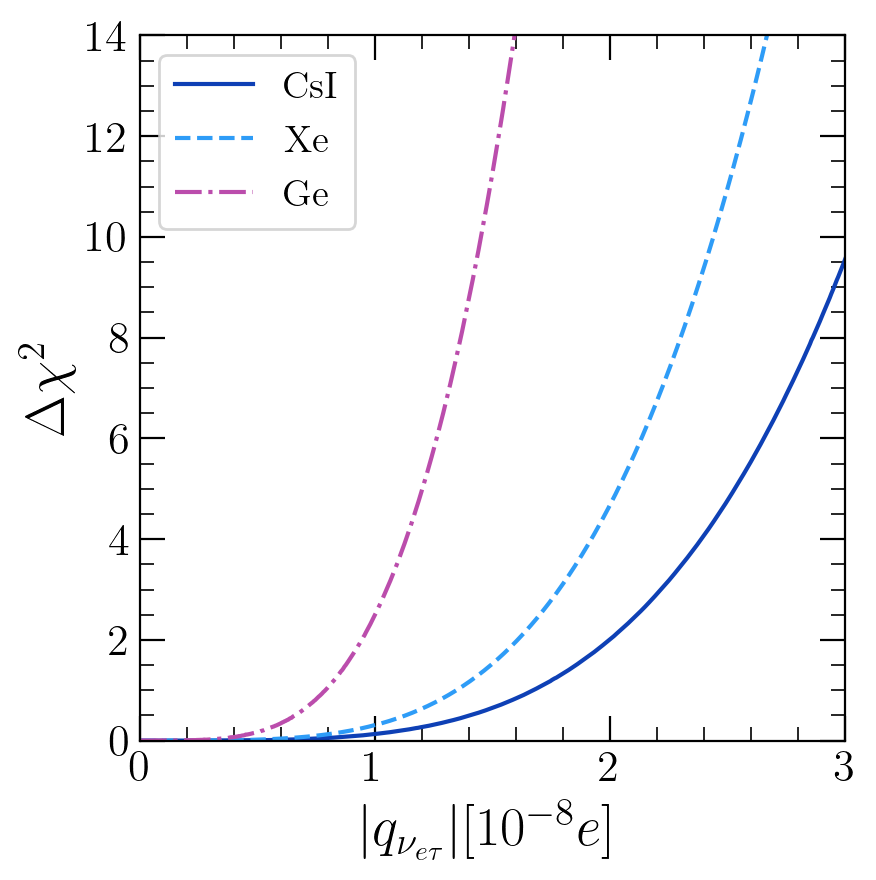}
\caption{Expected one dimensional $\Delta\chi^2$ profile for transition neutrino electric millicharges: $q_{\nu_{e\mu}}$ in the left panel, $q_{\nu_{\mu\tau}}$ in the central panel, and $q_{\nu_{e\tau}}$ in the right panel. We assume three years of data taking for three different detectors located at the ESS: CsI (solid blue), Xe (dashed light blue), and Ge (dash-dotted violet).  }
\label{fig:1D:results2}
\end{figure}

For completeness, we show in Fig. \ref{fig:1D:results2}, the results when considering individual transition neutrino charges, with $q_{\nu_{e\mu}}$ in the left panel, $q_{\nu_{\mu\tau}}$ in the central panel, and $q_{\nu_{e\tau}}$ in the right panel. We show in the figure the  $\Delta\chi^2$ profile for the three different detectors under the same color code as in Fig.~\ref{fig:1D:results}. Additionaly, we show in Table \ref{tab:qee} the expected bounds at a 90\% C.L. for each of the considered transition charges. Overall, we see that, again, the best constraint in all parameters is given by the Ge detector, reaching at most a sensitivity of $\sim 10^{-9}e$. For instance, in the case of Ge we obtain a expected sensitivity of $|q_{\nu_{e\mu}}| < 6.0\times 10^{-9}e$. We  can compare our results with those obtained through the combined analysis of COHERENT data in Ref.~\cite{Cadeddu:2020lky}, where it is obtained a current sensitivity for this parameter of $|q_{\nu_{e\mu}}| < 1.4\times 10^{-7}e$, showing that from the ESS it can be expected to lower the current CE$\upnu$NS bound by approximately two orders of magnitude.  

\subsection{Two neutrino millicharges}\label{sec:results:CEvNS}

\begin{figure}[t]
\centering
\includegraphics[width=0.48\textwidth]{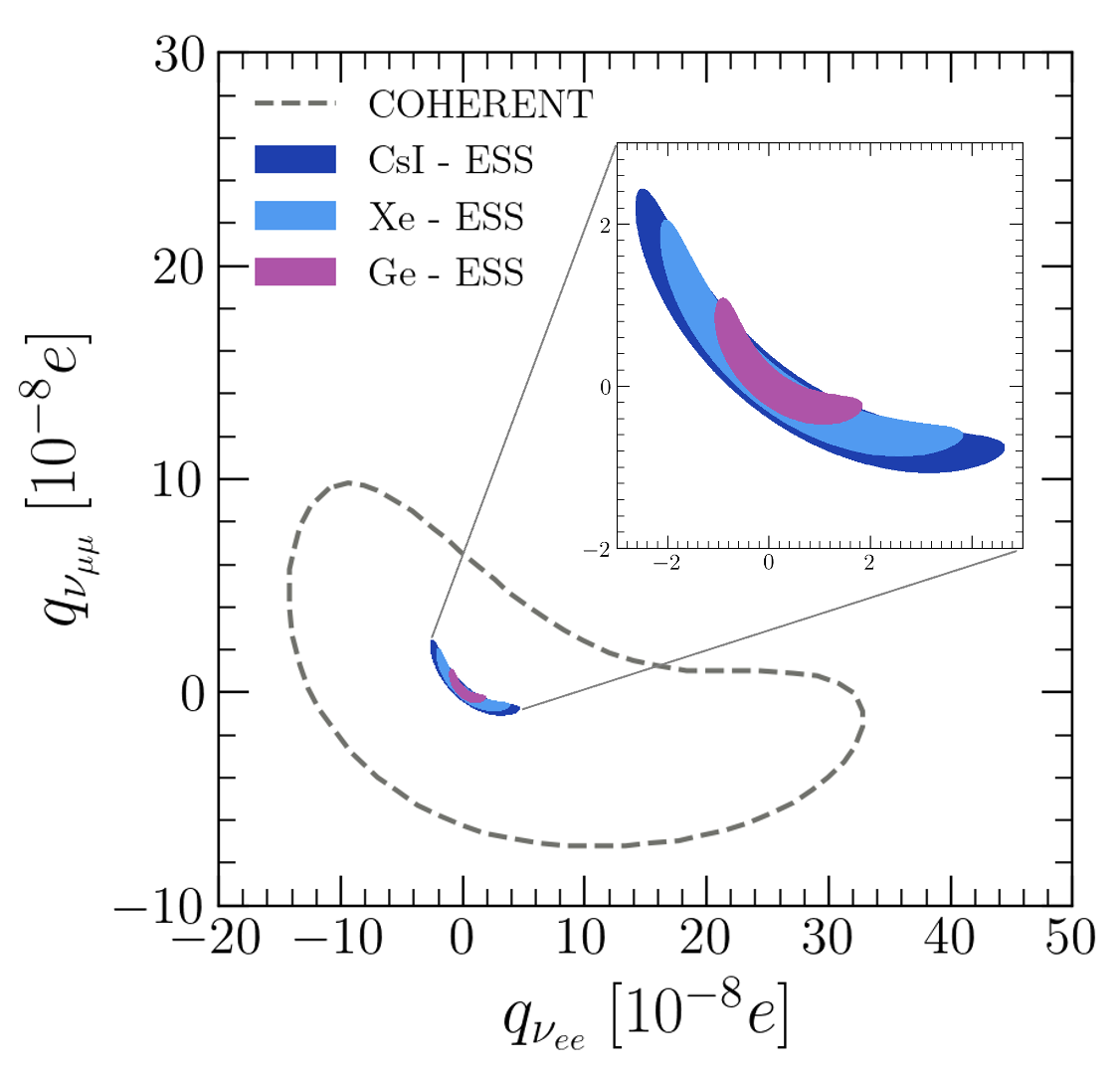}
\hspace{0.2cm}
\includegraphics[width=0.48\textwidth]{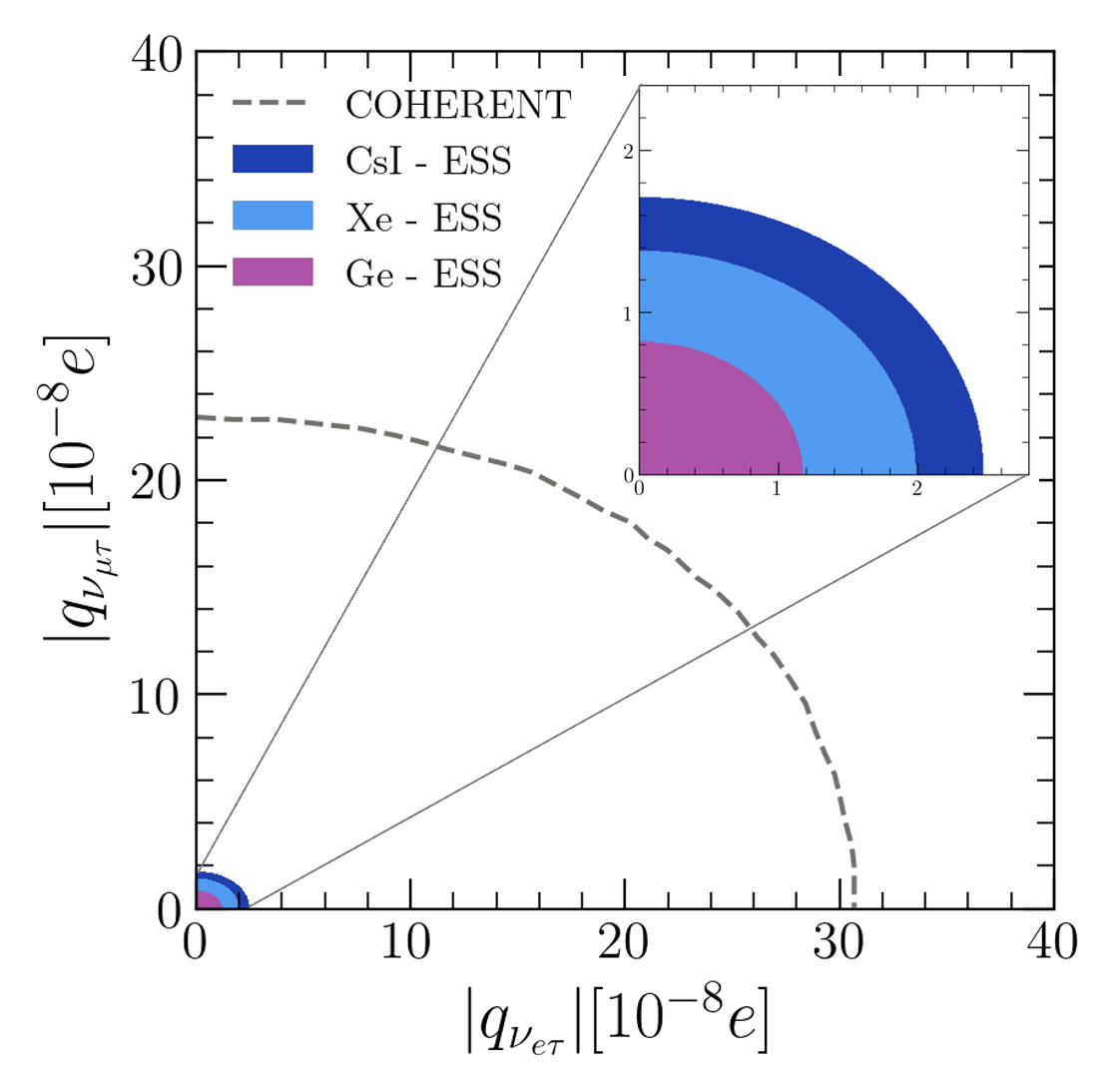}
\caption{Expected allowed regions, at a 90\% C.L., when allowing for two non-zero millicharges at a time. Left panel corresponds to the parameter space $(q_{\nu_{ee}}, q_{\nu_{\mu\mu}})$, and the right panel to the pair $(q_{\nu_{e\tau}}, q_{\nu_{\mu\tau}})$. We assume three years of data taking for three different detectors located at the ESS: CsI (dark blue), Xe (light blue), and Ge (violet). Dashed lines in both panels show the current bounds obtained through COHERENT data analysis, and are extracted from Refs.~\cite{DeRomeri:2022twg} and \cite{Cadeddu:2020lky} .   }
\label{fig:2D:results}
\end{figure}

We now study the sensitivity of the proposed ESS detectors to constrain millicharges when considering two of them to be non-zero at a time. The results of this analysis are given in Fig.~\ref{fig:2D:results}, where we show the expected sensitivity at a 90\% C.L. ($\Delta\chi ^2 < 4.61$) for the CsI (dark blue), Xe (light blue), and Ge (violet) detectors. The left panel of this figure, shows the results for the parameter space $(q_{\nu_{ee}}, q_{\nu_{\mu\mu}} )$, while the right panel shows the parameter space $(q_{\nu_{e\tau}}, q_{\nu_{\mu\tau}} )$. Again, we see that we can expect better constraints from the Ge detector, which is consistent with the results presented in the previous subsection. For comparison, we also show as dashed lines the current bounds obtained from the combination of the CsI and LAr COHERENT detectors. For the left panel, the combined COHERENT analysis is extracted from Ref. \cite{DeRomeri:2022twg}. We notice that the ESS can be expected to reduce the dimensions of the allowed parameter space by almost two orders of magnitude. Similarly, we compare the results in the right panel, with the COHERENT combined analysis adapted from Ref. \cite{Cadeddu:2020lky}. As in the previous case, we conclude that we can expect an improvement of around two orders of magnitude on the dimensions of the allowed parameter space. This latter result is of particular interest, since CE$\upnu$NS measurements can provide the only laboratory bounds for $q_{\nu_{\mu\mu}}$ and $q_{\nu_{\mu\tau}}$. This shows, complementary to the analysis presented in Ref.~\cite{Baxter:2019mcx}, the potential of the ESS to study electromagnetic properties of neutrinos. 

\subsection{Effects of electron scattering recoils}\label{sec:results:ES}

Now we turn our attention to the effect of considering electron scattering events for the study of neutrino millicharge in CE$\upnu$NS experiments. This is motivated by the fact that, for some detection technologies, it is not experimentally possible to distinguish between the signal produced by electron and nuclear recoils. For instance, the COHERENT LAr detector was able to distinguish between both recoils, while this was not the case for the CsI one used for the latest data release. Since the proposed CsI detector at the ESS shares the same target as COHERENT, we will particularly illustrate the effects of electron recoils for our considered CsI detector.

\begin{figure}[t]
\centering
\includegraphics[width=0.49\textwidth]{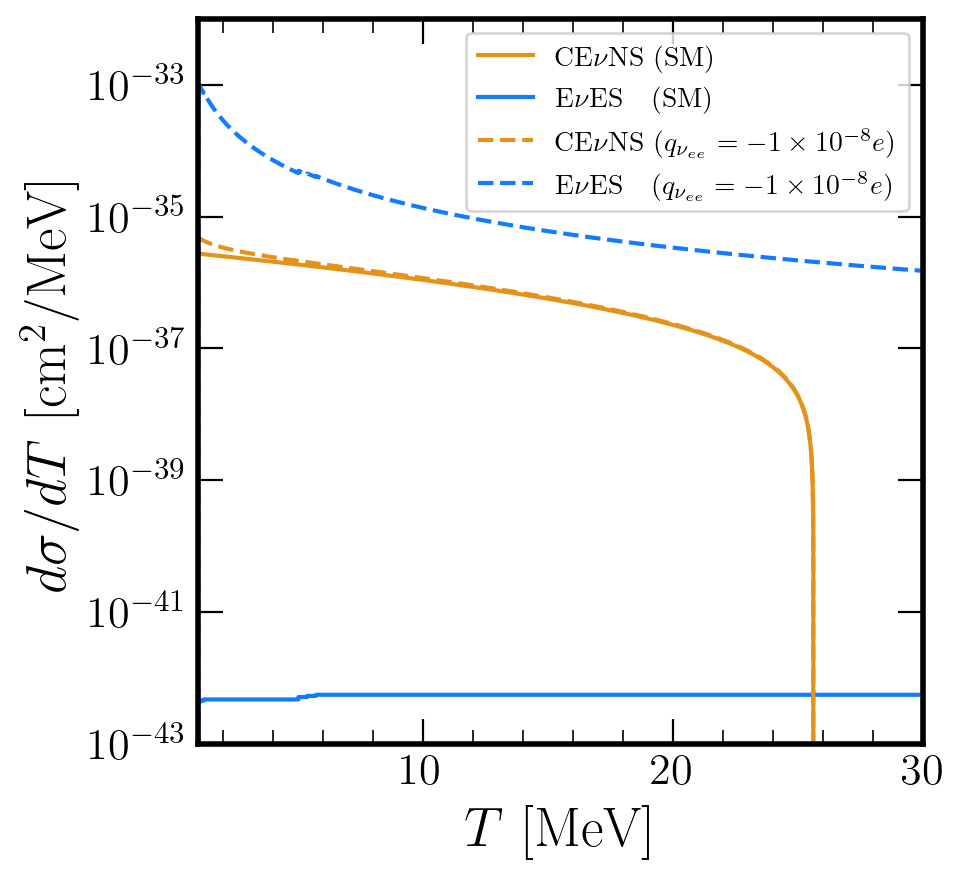}
\includegraphics[width=0.49\textwidth]{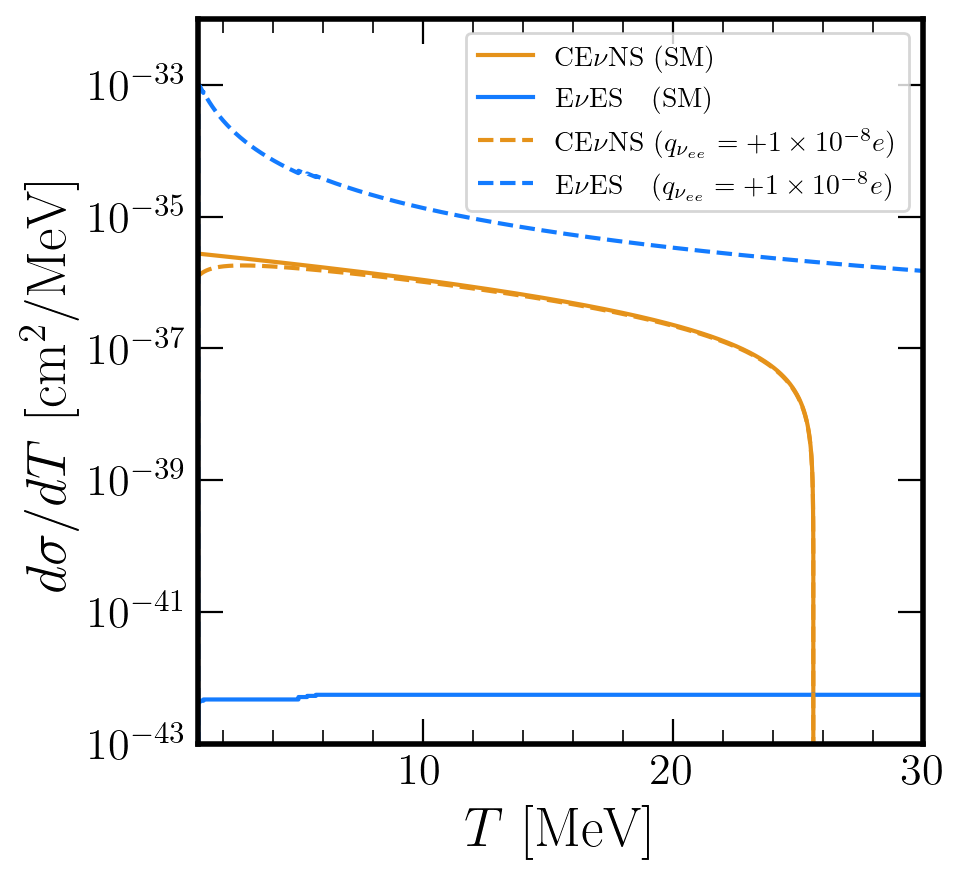}
\caption{CE$\upnu$NS and neutrino electron scattering (E$\upnu$ES) cross section comparison. Solid lines represent the SM prediction for CE$\upnu$NS (orange) and E$\upnu$ES (blue). Dashed lines show the corresponding predicted cross section assuming a non-zero value for $q_{\nu_{ee}}$. We consider $q_{\nu_{ee}} = -1\times10^{-8}$ in the left panel and $q_{\nu_{ee}} = +1\times10^{-8}$ in the right panel. In all cases, we assume a CsI target and an incoming neutrino energy of 10 MeV. }
\label{fig:cross:es:results}
\end{figure}

To understand why the addition of electron recoils is important for such detector, lets recall that, within the SM, the CE$\upnu$NS differential cross section is dominant with respect to the neutrino-electron elastic scattering (E$\upnu$ES) process. This can be seen from the solid lines in the two panels of Fig.~\ref{fig:cross:es:results}, where we show the SM cross-section for CE$\upnu$NS (orange) and electron scattering (blue) as given in Eqs.~\eqref{eq:cross} and \eqref{eq:cross:es}, respectively. To generate these figures, we have assumed a CsI detector and an incoming neutrino energy of 10~MeV. We notice that, for recoil energies of up to $\approx$~25~keV, the  CE$\upnu$NS SM cross section is around five orders of magnitude larger than the corresponding one for electron scattering. Then, we can safely say that, for the study of SM physics, the effects of electron scattering are not relevant when compared to nuclear recoils produced by CE$\upnu$NS interactions. However, the situation is different when considering some new physics scenarios as it is the case of a neutrino millicharge. To illustrate this, we show as dashed lines in the left panel of Fig. \ref{fig:cross:es:results} the CE$\upnu$NS (orange) and electron-scattering (blue) cross sections after considering $q_{\nu_{ee}} = -1\times10^{-8}e$. We clearly see that in both cases the cross section increases when compared to the corresponding SM prediction. However, for electron scattering the effect is such that now this is the dominant channel over the CE$\upnu$NS process. For completeness, we also show in the right panel of the same figure the comparison between the cross section for a positive value of $q_{\nu_{ee}} = +1\times10^{-8}e$. In this case, the electron-scattering cross section is even larger than the CE$\upnu$NS one in a sense that the CE$\upnu$NS cross section is lower than the corresponding SM prediction.

\begin{figure}[t]
\centering
\includegraphics[width=0.42\textwidth]{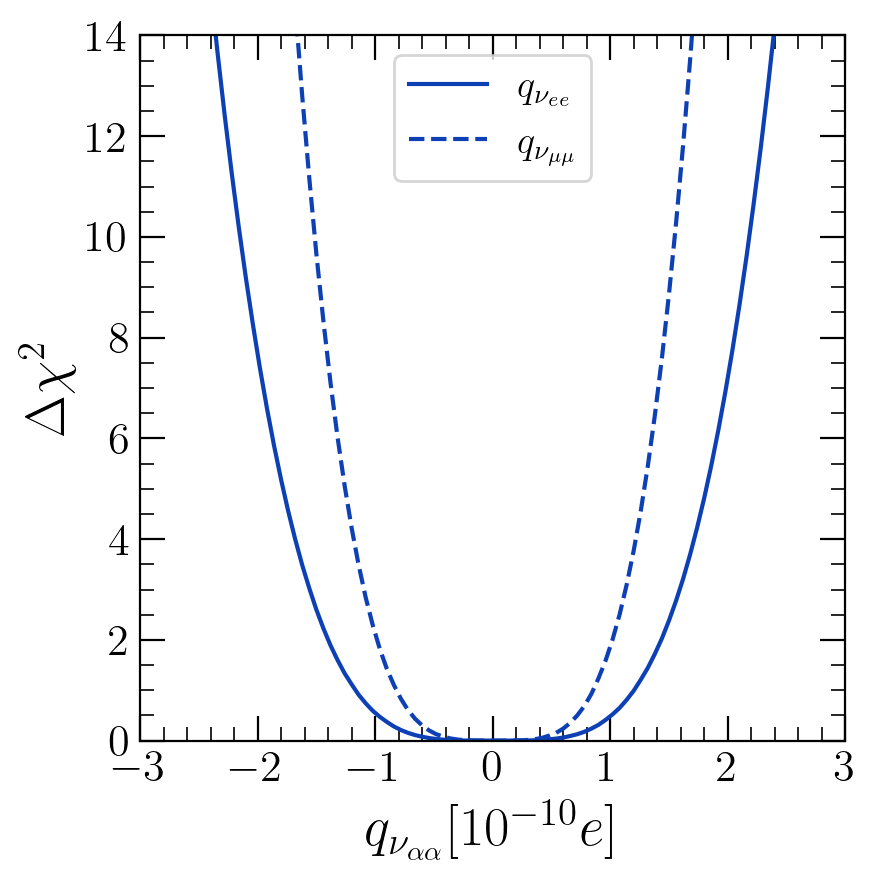}
\hspace{0.5cm}
\includegraphics[width=0.42\textwidth]{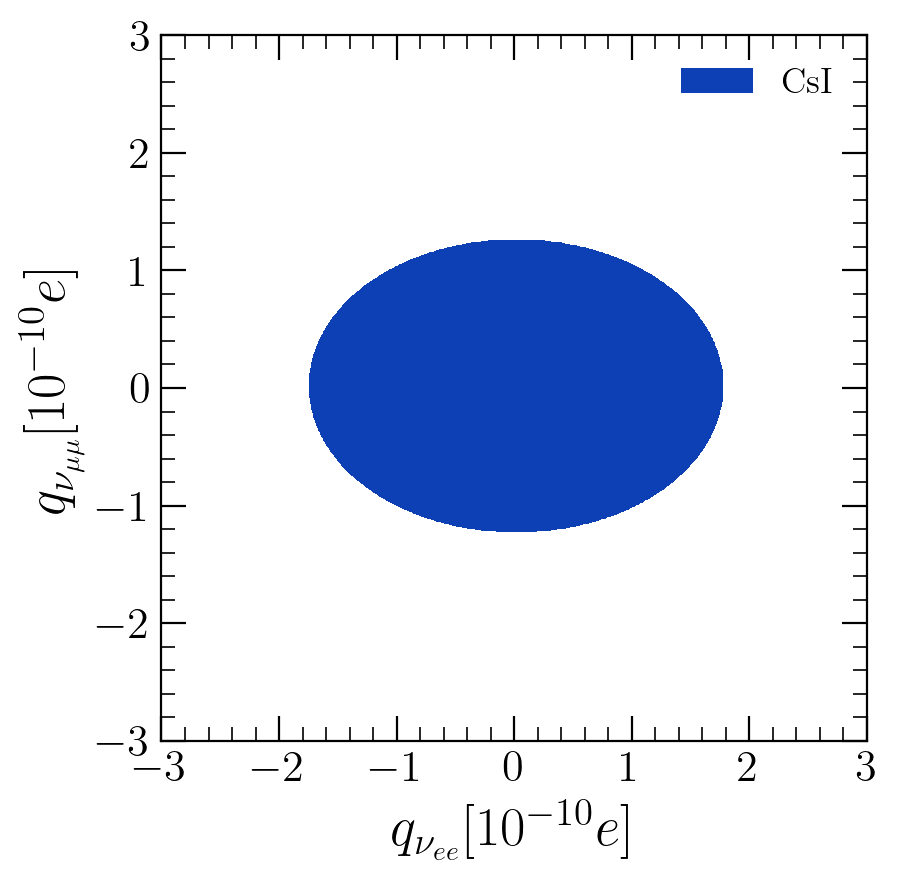}
\caption{Left panel: Expected one dimensional $\Delta\chi^2$ profile for $q_{\nu_{\alpha\alpha}}$ when considering CE$\upnu$NS and neutrino electron scattering interactions. Solid line indicates the results for $\alpha = e$ and dashed lines for $\alpha =\mu$. Right panel: Expected allowed region at a 90 \% C.L. for the parameter space $(q_{\nu_{ee}}, q_{\nu_{\mu\mu}})$, again considering CE$\upnu$NS and neutrino-electron scattering contributions. In both cases, we assume a CsI detector for three years of data taking at the ESS.  
 }
\label{fig:es:results}
\end{figure}

From the above discussion, we conclude that at low energy thresholds it is important to consider electron-scattering for our analysis. This statement can be generalized for any new physics scenario which contribution to the cross section includes an inverse dependence on the nuclear recoil energy, whether there is an interference with the SM cross section or not. For our case of interest, we show in the left panel of Fig.~\ref{fig:es:results} the results of the analysis including electron scattering events for the proposed CsI detector. The solid line in this figure corresponds to $q_{\nu_{ee}}$, while the dashed line to the case of $q_{\nu_{\mu\mu}}$. We see that at a 90\% C.L. ($\Delta\chi^2 = 2.71$), the expected bounds are of order $10^{-10}$, getting a result of $q_{\nu_{ee}} \in [-1.6, 1.6] \times 10^{-10}e$ and $q_{\nu_{\mu\mu}} \in [-1.1, 1.1] \times 10^{-10}e$.  This improves the current bound obtained in Ref. \cite{DeRomeri:2022twg}, when also electron scattering events are considered, by a factor of two for both cases. In addition, we show on the right panel of the same figure, the results at a 90\% C.L. when considering both millicharges to be different from zero at a time. We notice again an improvement of a factor of two on the dimensions of the allowed region when compared to the result obtained in Ref. \cite{DeRomeri:2022twg}. 

\section{Conclusions}\label{sec:conclusions}
We explored the sensitivity that three different detectors, to be located at the ESS, can reach to constrain electric neutrino millicharges. Particularly, we focused on the capability of a CsI, Xe, and Ge detectors, with similar characteristics as those currently under development for the study of CE$\upnu$NS interactions at the ESS. Our results show that the Ge detector will be more sensitive to test diagonal neutrino millicharges of order $1\times 10^{-9}e$, and of order $1\times 10^{-8}e$ for the transition case.
This sensitivity would improve the existing bounds obtained from current CE$\upnu$NS dedicated  experiments by two orders of magnitude. In addition, we found that, by including neutrino-electron scattering interactions for the CsI detector,  neutrino millicharges can be constrained for values of up to order $10^{-10}e$.  Our results show that CE$\upnu$NS will still not be competitive with reactor bounds as that from GEMMA. However, it shows a major improvement with respect to current CE$\upnu$NS experiments for the study of electromagnetic properties of neutrinos through different interaction channels.

\acknowledgements 
\noindent This work has been supported by the Spanish grants PID2020-113775GB-I00 (MCIN/AEI/ 10.13039/501100011033) and CIPROM/2021/054 (Generalitat Valenciana). G.S.G. acknowledges financial support by the Grant No. CIAPOS/2022/254 (Generalitat Valenciana). A. Parada thanks Escuela Superior de Administración Pública (ESAP) for the support through research project E2\_2024\_09.  

\bibliographystyle{utphys}
\bibliography{references} 

\end{document}